\documentclass[aps,prb,twocolumn,showpacs,preprintnumbers,amsmath,amssymb,superscriptaddress,floatfix]{revtex4}
\usepackage{graphicx}
\usepackage{dcolumn}
\usepackage{bm}
\begin{document}
\renewcommand{\thefigure}{\arabic{figure}}
\title{Effective mass suppression in
a ferromagnetic two-dimensional electron liquid}
\author{Reza Asgari}
\affiliation{School of Physics, Institute for Research in Fundamental
Sciences, (IPM) 19395-5531 Tehran, Iran}
\author{T. Gokmen}
\affiliation{Department of Electrical Engineering, Princeton University,
Princeton, New Jersey 08544, USA}
\author{B. Tanatar}
\affiliation{Department of Physics, Bilkent University,
Bilkent, Ankara 06800, Turkey}
\author{Medini Padmanabhan}
\affiliation{Department of Electrical Engineering, Princeton
University, Princeton, New Jersey 08544, USA}
\author{M. Shayegan}
\affiliation{Department of Electrical Engineering, Princeton University, Princeton, New Jersey 08544, USA}

\begin{abstract}
We present numerical calculations of the electron effective mass in
an interacting, ferromagnetic, two-dimensional electron system. We
consider quantum interaction effects associated with the
charge-density fluctuation induced many-body vertex corrections. Our
theory, which is free of adjustable parameters, reveals that the
effective mass is suppressed (relative to its band value) in the
strong coupling limit, in good agreement with recent
experimental results.
\end{abstract}
\maketitle

\section{Introduction}
Two-dimensional electron systems
(2DESs) realized at semiconductor interfaces are of continuing
interest \cite{ando,gv_book} from both basic physics and
technological points of view.
As a function of the interaction strength, which is characterized
by the ratio $r_s$ of the Coulomb energy to Fermi energy, many novel
correlated ground states have been predicted such as a paramagnetic
liquid ($r_s < 26$), ferromagnetic liquid ($26 < r_s < 35$) and
Wigner crystal ($r_s > 35$) \cite{AttaccalitePRL02}.
In the paramagnetic liquid
phase, interaction typically leads to an enhancement of effective
mass ($m^*$) and spin susceptibility (${\chi}^*$ ${\propto}$ $g^{*}
m^{*}$), where $g^*$ is the Land{\'e} $g^*$-factor. Effective mass
is an important concept in Landau's Fermi liquid theory since it
provides a direct measure of the many-body interactions in the
electron system as characterized by increasing $r_s$.

The effective mass $m^{*}$ renormalized by interactions has been
experimentally studied
\cite{SmithPRL72,PanPRB99,ShashkinPRL03,TanPRL05,PadmanabhanPRL08}
for various paramagnetic 2DESs as a function of $r_{s}$. In the
highly interacting, dilute, paramagnetic regime ($3 < r_{s} < 26$),
$m^{*}$ is typically significantly enhanced compared to its band
value, $m_{b}$, and tends to increase with increasing
$r_{s}$~\cite{SmithPRL72,PanPRB99,ShashkinPRL03,TanPRL05,
PadmanabhanPRL08,KwonPRB94,AsgariSSC04,AsgariPRB05,AsgariPRB06,
GangadharaiahPRL05,ZhangPRL05}. A question of particular interest is
the dependence of $m^{*}$ on the 2D electrons' spin and valley
degrees of freedom as these affect the exchange interaction. Recent
measurements of $m^{*}$ for 2D electrons confined to AlAs quantum
wells revealed that, when the 2DES is fully valley- and
spin-polarized, $m^{*}$ is {\it suppressed} down to values near or
even slightly below $m_b$
\cite{PadmanabhanPRL08,GokmenPRL08,GokmenUNP09}. Note that in these
experiments, $r_s < 22$ so that the 2DES is in the paramagnetic
regime, but a strong magnetic field is applied in order to fully
spin-polarize the electrons. Here we present theoretical
calculations indicating that the $m^{*}$ suppression is caused by
the absence (freezing out) of the spin fluctuations. The results of
our $m^{*}$ calculations are indeed in semi-quantitative agreement
with the measurements.

Previous theoretical calculations of the effective mass are mostly
performed within the framework of Landau's Fermi liquid theory whose
key ingredient is the quasiparticle (QP) concept and its
interactions. This entails the calculation of effective
electron-electron interactions which enter the many-body formalism
allowing the calculation of effective mass. A number of works
considered different variants of the leading order in the screened
interaction for the
self-energy~\cite{em,yarlagadda_1994_2,em_bohm,em_dassarma,DasPRB04,zhang,
AsgariSSC04,AsgariPRB05,AsgariPRB06} from which density,
spin-polarization, and temperature dependence of effective mass are
obtained. In these calculations the on-shell
approximation~\cite{em_bohm,em_dassarma,DasPRB04} yields a diverging
effective mass but the full solution of the Dyson equation yields
only a mild enhancement.~\cite{AsgariSSC04,AsgariPRB05,AsgariPRB06}
Almost all these works considered a paramagnetic 2DES as past
experiments concentrated on the effective mass enhancement in
partially spin polarized 2D systems with $r_s < 26$.

\section{Theory}
We consider a ferromagnetic 2DES as a model
for a system of electronic carriers with band mass $m_b$ in a semiconductor
heterostructure with dielectric constant $\kappa$. The bare
electron-electron interaction is given by $v_{\bf q}=2\pi
e^2/(\kappa q)$. At zero temperature there is only one relevant
parameter for the homogeneous, ferromagnetic 2DES, the usual
Wigner-Seitz density parameter $r_s=(\pi n a^2_B)^{-1/2}$ in which
$a_B=\hbar^2 \kappa/(m_b e^2)$ is the Bohr radius in the medium of
interest.

\begin{figure}[ht]
\label{f1}
\begin{center}
\includegraphics[scale=0.47]{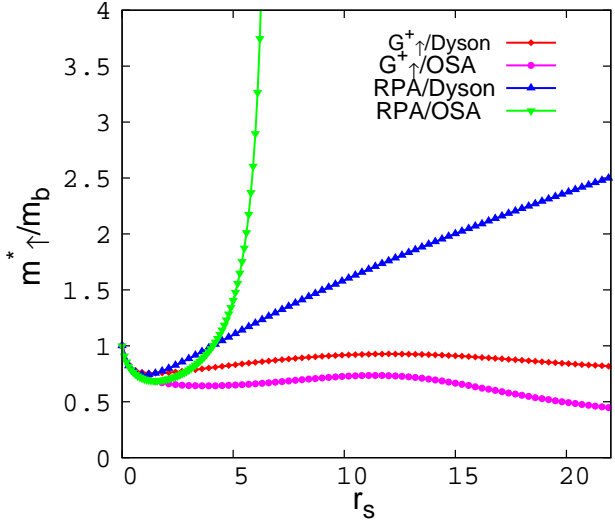}
\caption{(color online). Many-body effective mass as a function of $r_s$ for
$0\leq r_s\leq 22$ for a ferromagnetic 2DES.}
\end{center}
\end{figure}

The QP self-energy with momentum $\bf k$ and frequency $\omega$ in a
fully polarized electron system can be written as
\begin{widetext}
\begin{equation}\label{self}
\Sigma^{\uparrow}({\bf k},\omega)=
-\int \frac{d^2 {\bf q}}{i (2\pi)^2}v_{\bf q}\int_{-\infty}^{\infty}
\frac{d\Omega}{2\pi}\frac{1}{\varepsilon({\bf q}, \Omega)}\,\left [\frac{1-n_{\rm F}(\xi^{\uparrow}_{\bf k})}{\omega+ \Omega-\xi^{\uparrow}_{{\bf k}+{\bf q}}/\hbar+i\eta}+ \frac{n_{\rm F}(\xi^{\uparrow}_{\bf k})}{\omega+ \Omega-\xi^{\uparrow}_{{\bf k}+{\bf q}}/\hbar-i\eta}\right ]\,.
\end{equation}
\end{widetext}
Here $\xi^{\uparrow}_{\bf k}=\varepsilon_{\bf k}-\varepsilon_F$
where $\varepsilon_{\bf k}=\hbar^2{\bf k}^2/(2m_b)$ is the
single-particle energy with
$\varepsilon_F=\hbar^2{k^{\uparrow}_F}^2/(2m_b)$ and
$k^{\uparrow}_F=(4\pi n_{\rm \scriptscriptstyle 2D})^{1/2}=2/(r_s
a_B)$, respectively, being the Fermi energy and wave vector; $n_{\rm
F}(k)$ is the Fermi function. In Eq.~(\ref{self}), $\varepsilon({\bf
q},\omega)$ is the dynamical screening function for which we use the
form appropriate for a ferromagnetic 2DES derived
from Kukkonen-Overhauser effective interaction.~\cite{kukkonen}
The many-body exchange and correlation (XC) effects are introduced
through the
local-field factors (LFF) $G_{\sigma,\sigma'}(q,\omega)$
($\sigma$ and $\sigma'$ are spin indices) take the
Pauli-Coulomb hole around a charged particle into account.
The dynamical screening function reads
\begin{equation}\label{epsilon}
\frac{1}{\varepsilon({\bf q},\omega)}=
1+v_{\bf q}\,\left[1-G^{+}_{\uparrow}({\bf q},\omega)\right]^2\,
\chi_{\rm \scriptstyle C}({\bf q},\omega)~,
\end{equation}
where $G^+_\uparrow$ is the LFF associated with
charge-fluctuations.
This expression is similar to the Kukkonen and Overhauser
interaction~\cite{kukkonen} where the spin-fluctuation term is
dropped. A similar expression has also been reported in
Refs.\,[\onlinecite{ng,tanaka}]. In Eq.\,(2) $\chi_{\rm \scriptstyle
C}({\bf q},\omega)$ represents the density-density response
function, which in turn is determined by the local-field factor
$G^{+}_{\uparrow}({\bf q},\omega)$ via the relation
\begin{equation}\label{cc}
\chi_{\rm \scriptstyle C}({\bf q},\omega)
=\frac{\chi^0_{\uparrow}({\bf q},\omega)} {1-v_{\bf
q}[1-G^+_{\uparrow}({\bf q},\omega)] \chi^0_{\uparrow}({\bf
q},\omega)}\, ,
\end{equation}
in which $\chi^0_\uparrow(q,\omega)$ is the density response function
of the spin-polarized electrons. The expression for the noninteracting
density response function on the imaginary frequency axis is obtained
for use in Eq.~(\ref{cc}) as
\begin{equation}
\chi^0_{\uparrow}({\bf q},i\Omega)=\frac{m_b^2}{2 \pi \hbar^2 q^2}
\left(\sqrt{2}\sqrt{a_{\uparrow}+\sqrt{a_{\uparrow}^2+
\left(\frac{q^2 \Omega}{\hbar m_b}\right)^2}}-\frac{q^2}{m_b}\right)\, ,
\end{equation}
where we have defined $a_{\uparrow}=q^4/4m_b^2-q^2 {k^{\uparrow}_F}^2/m_b^2
-\Omega^2/\hbar^2$. It is evident that setting $G^+_{\uparrow}({\bf q},\omega)=0$,
we recover the standard random phase approximation (RPA). In what follows,
we shall make the common approximation of neglecting the
frequency dependence of $G^+_\uparrow$.

Quite generally, once the QP retarded self-energy is known, the QP
excitation energy $\delta{\mathcal E}^{\uparrow}_{\rm QP}({\bf k})$,
which is the QP energy measured from the chemical potential
$\mu^{\uparrow}$ of the interacting ferromagnetic
2DES, can be calculated by solving self-consistently the Dyson
equation
\begin{equation}
\delta {\mathcal E}^{\uparrow}_{\rm QP}({\bf
k})=\xi^{\uparrow}_{\bf k}+ \left. \Re e \Sigma^{R}_{\rm
\scriptstyle ret}({\bf k},\omega)\right|_ {\omega=\,\delta {\mathcal
E}^{\uparrow}_{\rm QP}({\bf k})/\hbar}\, .
\end{equation}
Alternatively, the QP excitation energy can also be calculated from
\begin{equation}
\delta {\mathcal E}^{\uparrow}_{\rm
QP}({\bf k})= \xi^{\uparrow}_{\bf k}+\left. \Re e
\Sigma^{R}_{ret}({\bf k},\omega)\right|_
{\omega=\xi^{\uparrow}_{\bf k}/\hbar}\, .
\end{equation}
This is called the on-shell approximation (OSA) and it is argued~\cite{rice}
to be a better approach than solving the full Dyson equation since
noninteracting Green function is used in Eq.\,(1). Here $\Re e
\Sigma^{R}_{ret}({\bf k},\omega)$ is defined as $\Re e
\Sigma^{\uparrow}_{ret}({\bf k},\omega)-
\Sigma^{\uparrow}_{ret}({\bf k^{\uparrow}_F},0)$.

The effective mass $m^*_{\uparrow}(k)$ is now calculated from
\begin{equation}\label{ms}
\frac{1}{m^*_{\uparrow}(k)}=\frac{1}{\hbar^2 k}
\frac{d \delta {\mathcal E}^{\uparrow}_{\rm QP}(k)}{dk}\,,
\end{equation}
where for $\delta{\mathcal E}^\uparrow_{\rm QP}$ we have at our
disposal the Dyson and OSA approaches. Evaluating
$m^*_{\uparrow}(k)$ at $k=k^{\uparrow}_F$, one gets the QP effective
mass at the Fermi contour. Clearly from Eqs.\,(\ref{epsilon}) and
(\ref{cc}) LFF is the basic quantity for an
evaluation of the QP properties. We have used the parameterized
forms of LFFs $G^+(q,\zeta)$ and $G^{-}(q,\zeta)$ (and in particular
$G^+_\uparrow(q)=G^+(q,\zeta=1)$ where $\zeta$ is the spin
polarization) of Moreno and Marinescu.~\cite{moreno}

\section{Results and discussion}
We now present our numerical results,
which are based on the LFF $G^+_\uparrow(q)$ as input.
In Fig.~1 we show our
numerical results of the QP effective mass both in OSA and Dyson
approximations. The QP effective mass suppression is substantially
smaller in the Dyson equation calculation than in the OSA; the
reason is that a significant cancellation occurs between the
numerator and the denominator in the effective mass expression in
the Dyson approach. To clarify the effect of charge-density
fluctuation we have also shown the RPA results which do not take
the strong many-body fluctuations into account. Note that the LFF
takes into account multiple scattering events to infinite order
as compared to the RPA where these effects are neglected. In the limit of
small $\zeta$ and $r_s\rightarrow 0$, the effective mass can be
analytically shown to be
$m^*_{\uparrow}/m_b=1+(1-\zeta/2.0)r_s\ln r_s/(\sqrt{2} \pi)$
which our numerical calculations faithfully reproduce.

\begin{figure}[ht]
\label{f2}
\begin{center}
\includegraphics[scale=0.47]{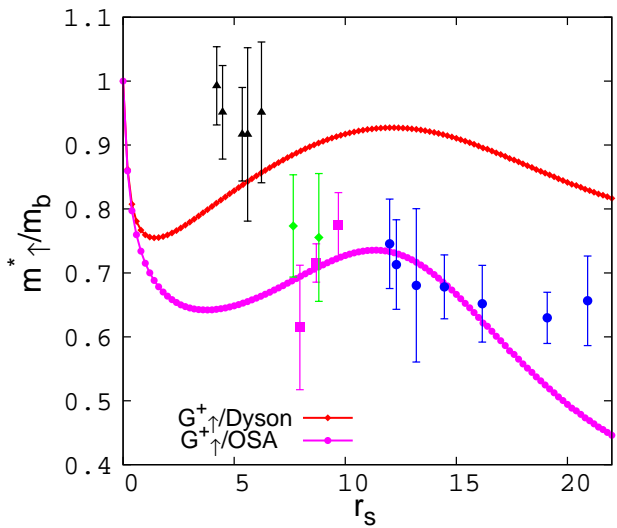}
\caption{(color online). Many-body effective mass as a function of
$r_s$ for $0\leq r_s\leq 22$ for the ferromagnetic 2DES in
comparison to experiments
in Ref.\,[\onlinecite{PadmanabhanPRL08,GokmenUNP09}]. Different
symbols denote different samples; triangles: A, squares: B, circles: C,
and diamonds: D.}
\end{center}
\end{figure}

In Fig.~2 we compare our effective mass calculations with the
experimental results
\cite{PadmanabhanPRL08,GokmenPRL08,GokmenUNP09}. The measurements
were made on 2DESs confined to modulation-doped AlAs quantum wells
(QWs) of width 4.5, 11, 12, and 15\,nm (samples A, B, C, and D).
These samples were grown on GaAs substrates using molecular beam
epitaxy. In bulk AlAs, electrons occupy three degenerate ellipsoidal
conduction band valleys at the X-points of the Brillouin zone with
longitudinal and transverse effective masses $m_l$=1.05 and
$m_t$=0.205 (in units of the free electron mass). Thanks to the
slightly larger lattice constant of AlAs compared to GaAs, the AlAs
QW layer is under bi-axial compressive strain. Because of this
compression, the 2DES in the wider QW samples (B, C, and D) occupy
two in-plane valleys with their major axes lying in the plane
\cite{ShayeganPSS06}. In our measurements on these samples, we
applied uni-axial, in-plane strain to break the symmetry between
these two valleys so that only one in-plane valley, with an
$anisotropic$ Fermi contour and band effective mass of $m_b =
\sqrt{m_l m_t} = 0.46$ is occupied \cite{ShayeganPSS06}. In sample
A, however, thanks to its very small QW width, the confinement
energy of the out-of-plane valley is lower (because of its larger
mass along the growth direction), so that the electrons occupy this
valley and therefore have an $isotropic$ Fermi contour and band
effective mass is $m_b = m_t = 0.205$ \cite{ShayeganPSS06}. The
effective masses were deduced from the temperature dependence of the
Shubnikov-de Haas oscillations, the details of which are given in
Refs. \cite{PadmanabhanPRL08,GokmenPRL08,GokmenUNP09}. We emphasize
that the data shown here (Fig.\,2) were taken on single-valley 2DESs
which were subjected to sufficiently large magnetic fields to fully
spin polarize the electrons .

It appears in Fig.\,2 that the OSA accounts overall for the observed
reduction of $m^*_\uparrow$ below the band value reasonably well.
The agreement is particularly good for the wider samples (B, C, and
D) which have $r_s>7$. The $m^*$ data for the narrowest sample (A),
however, fall above the theoretical predictions. We do not know the
reason for this discrepancy. However, we point out that, besides the
difference in the shapes of the Fermi contour, there is another
difference between sample A and the other three samples. Because of
the very narrow width of sample A's quantum well and the prevalence
of interface roughness scattering \cite{VakiliAPL06}, the mobility
of the electrons in this sample is much lower (about a factor of 6)
than in other samples for comparable $r_s$. It is possible that the
higher disorder in sample A is responsible for $m^*$ being larger; this
conjecture is indeed consistent with the results of calculations
\cite{AsgariSSC04} which predict a larger $m^*$ for
more disordered samples.

From Figs.\,1 and 2 we draw two main conclusions. (i) The RPA and
present results are rather similar in the weak coupling limit
($r_s<1$). (ii) In the strong coupling regime ($r_s > 3$), however,
our theoretical calculations which incorporate the proper many-body
effects exhibit a mass suppression, similar to the experimental
data, while the RPA results show a mass enhancement and are far from
the experimental data. We emphasize that the effective mass at the
Fermi contour is significantly suppressed in the fully polarized
case because of the absence of spin-fluctuation contribution. This
suppression suggests that the anti-symmetric Landau parameter $F^a_1
<0$ and thus higher angular momentum Landau parameters may be
negligible in a fully spin-polarized 2DES.

\begin{figure}[ht]
\label{f3}
\begin{center}
\includegraphics[scale=0.47]{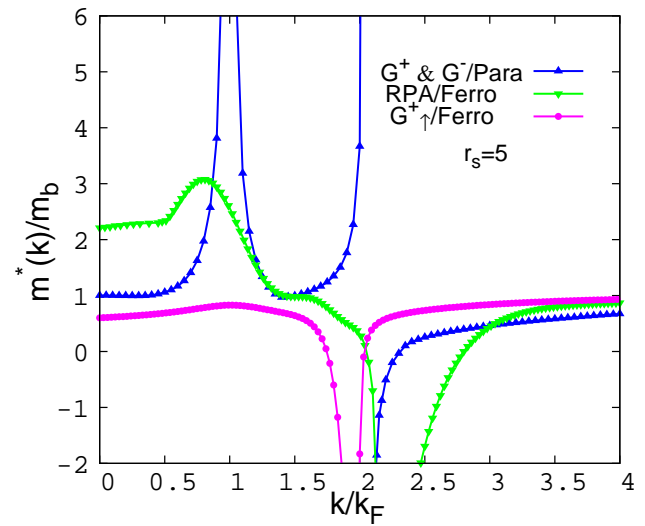}
\caption{(color online). Many-body on-shell effective mass as a function of
$k/k_F$ at $r_s =5$ for 2DES with the combined effect
of charge fluctuations in comparison to paramagnetic 2DES.}
\end{center}
\end{figure}

\begin{figure}[ht]
\label{f4}
\begin{center}
\includegraphics[scale=0.47]{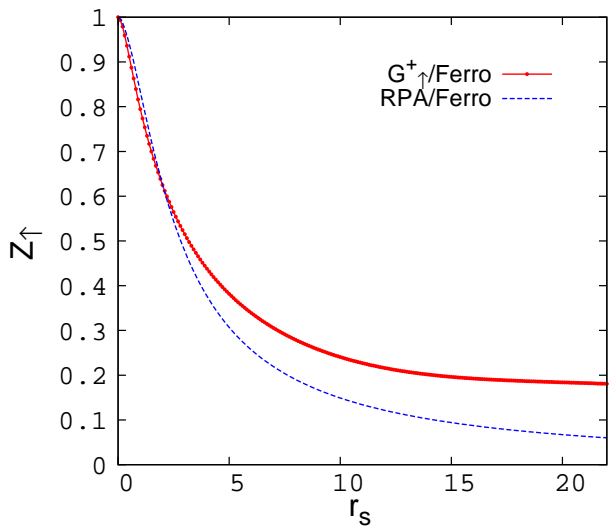}
\caption{(color online). Renormalization constant $Z_{\uparrow}$
as a function of $r_s$ for
$0 <r_s < 22$ for a ferromagnetic 2DES.}
\end{center}
\end{figure}

To gain further insight to the density dependence of $m^\ast$,
we have calculated the on-shell effective mass as a function of
particle momentum $k$ using Eq.~(\ref{ms}) evaluated at
$\omega(k)=\xi^{\uparrow}_{\bf k}/\hbar$ and $r_s= 5$. More
specifically, we use
\begin{equation}\label{mass_osa}
\frac{m_b}{m^*_{{\uparrow}}(k)}= 1+\frac{m_b}{\hbar^{2} k}
\frac {d}{dk} \Re e \Sigma^{\uparrow}_{\rm ret}
(k,\xi^{\uparrow}_k),
\end{equation}
for a ferromagnetic case.
The results for both paramagnetic and ferromagnetic cases are shown
in Fig.~3. $m^*(k)$ for a paramagnetic 2DES by using
$G^+(q,\zeta=0)$ and $G^-(q,\zeta=0)$ has a sharp peak around
$k\approx k_{\rm F}$ where $k_{\rm F}=2/(r_s a_B)$ and a resonance
like divergent behavior around $k\approx 2k_{\rm F}$. The peak
around $k_{\rm F}$ is associated with spin fluctuations and the
divergent behavior around $2k_{\rm F}$ is related to density
fluctuations.~\cite{ng,zhang} In particular, the latter divergence
has been extensively studied by Zhang {\it et al}.~\cite{zhang}
within the RPA. It is related to the dispersion instability and
coincides with the plasmon emission. $m^*_\uparrow(k)$ for the
ferromagnetic 2DES, on the other hand, clearly shows the
disappearance of the peak associated with spin fluctuations. Thus,
$m^\ast_\uparrow(k)$ is very weakly momentum dependent for
$k<k^{\uparrow}_F$, since there is a substantial cancelation between
the residue and the exchange plus line self-energy contribution in
this regime which make the real part of the retarded self-energy
approximately linear with respect to $k$.~\cite{AsgariPRB05} The
divergence associated with charge fluctuations is still present,
showing a negative peak around $k=2k_{\rm F}$. $m^\ast_\uparrow(k)$
calculated within the RPA reproduces quantitatively the divergent
behavior associated with charge fluctuations but shows some
structure for $k\leq k^\uparrow_{\rm F}$, therefore failing to
account for the absence of spin fluctuations. Our density-dependent
effective mass results (Figs.\,1 and 2) are consistent with
$m^\ast(k)$ calculations which we have checked for a range of $r_s$
values.

We have also calculated the
renormalization factor $Z_{\uparrow}(r_s)$ which is equal to the
discontinuity in the momentum distribution at $k_F$ and defined by
$Z^{-1}_{\uparrow}=1-\hbar^{-1} \left.\partial_{\omega}
\Re e \Sigma^{\uparrow}_{\rm ret}(k,\omega)\right|_{k=k^{\uparrow}_F,\omega=0}$.
The effect of charge-fluctuations is to make the $Z_{\uparrow}$ values
larger at large $r_s$ compared to the case when they are not
included as shown in Fig.~4. This means that charge-density
fluctuations tend to stabilize the system, whereas the RPA works
in the opposite direction.\cite{AsgariPRB05} In the present case
including the LFF helps preserve the Fermi liquid picture
in the low density regime.

We have performed our numerical calculations for strictly 2DES.
As indicated above, experimental samples have a finite thickness
in the range of $5-15$\,nm.
Our theoretical model may be extended to include the finite quantum
well width effects in the following manner. Choosing, say, an infinite
square well model with width $L$ will modify the bare Coulomb interaction,
$v_{\bf q}\rightarrow v_{\bf q}F(qL)$, in which $F(x)$ is a form
factor.\cite{gold} For consistency, one should also calculate
the local-field factor $G^+_\uparrow(q)$ using the same model
for the finite width effects. This would provide a better comparison
with experiments. In our case, the local-field factor we use
was constructed\cite{moreno} by the quantum Monte Carlo data for
a strictly 2DES and it is not straightforward to incorporate the finite with
effects within such an approach. Previous calculations\cite{AsgariPRB06}
of the effective mass for a paramagnetic 2DES suggest that the effect
of a finite thickness is to suppress $m^*$. Therefore, we surmise that
a similar qualitative effect would occur for the ferromagnetic 2DES.
On the other hand, the finite temperature and disorder effects
have a tendency to enhance the effective mass\cite{DasPRB04,AsgariSSC04} which may
lead to a cancellation. These issues require a more systematic study.

\section{Summary}
In conclusion, our theoretical calculations incorporating
the proper Pauli-Coulomb hole and multi-scattering processes
show that in an interacting, fully spin-polarized 2DES the absence of spin
fluctuations reduces the effective mass below its band value, in
agreement with experimental data. Our results also demonstrate the
inadequacy of RPA to account for the observed effective mass suppression.

\acknowledgments{R.\,A. thanks M. Polini for helpful
discussions. The work at Princeton University
was supported by the NSF. B.\,T. is supported by TUBITAK (No.
108T743) and TUBA.}

\end{document}